\documentclass{iau}

\newcommand{\gcmt}{{\rm g/cm}^3}

\newcommand{\lsim}{\stackrel{\textstyle <}{_\sim}}
\newcommand{\gsim}{\stackrel{\textstyle >}{_\sim}}
\newcommand{\msun}{M_{\odot}}

\newcommand{\fm}{{\rm fm}}
\newcommand{\ergs}{{\rm erg/s}}

\usepackage{graphicx} 

\title[IAUS291.~~Structure of Quark Stars] 
{Structure of Quark Stars} 

\author[One \& Two]  
{Fridolin Weber$^1$,  
  Milva Orsaria$^2$, \thanks{Home address: CONICET, Rivadavia 1917,
    1033 Buenos Aires; Gravitation, Astrophysics and Cosmology Group,
    Facultad de Ciencias Astron{\'o}micas y Geofisicas, Paseo del
    Bosque S/N (1900), Universidad Nacional de La Plata UNLP, La
    Plata, Argentina.}
  Hilario Rodrigues$^3$ \thanks{Home address: Centro Federal de
    Educa\c{c}\~ao Tecnol\'{o}gica do Rio de Janeiro, Av
    Maracan$\tilde{a}$ 249, 20271-110, Rio de Janeiro, RJ, Brazil.}
  \and Shu-Hua Yang$^4$ \thanks{Home address: Institute of
    Astrophysics, Huazhong Normal University, Wuhan, 430079,
    P. R. China.} }
\affiliation{Department of Physics, San Diego State University, 5500
  Campanile Drive, San Diego,
  California 92182, USA\\
  $^1$ email: {\tt fweber@mail.sdsu.edu} \\[\affilskip]
  $^2$ email: {\tt morsaria@fcaglp.fcaglp.unlp.edu.ar} \\[\affilskip]
  $^3$ email: {\tt harg@cefet-rj.br}\\[\affilskip]
  $^4$ email: {\tt ysh@phy.ccnu.edu.cn}\\[\affilskip]
}
\pubyear{2012}
\volume{291}  
\jname{\mbox{Neutron Stars and Pulsars: Challenges and Opportunities
    after 80 years}} \editors{J. van Leeuwen, ed.}
\begin{document}

\maketitle

\begin{abstract}
  This paper gives an brief overview of the structure of hypothetical
  strange quarks stars (quark stars, for short), which are made of
  absolutely stable 3-flavor strange quark matter. Such objects can be
  either bare or enveloped in thin nuclear crusts, which consist of
  heavy ions immersed in an electron gas.  In contrast to neutron
  stars, the structure of quark stars is determined by two (rather
  than one) parameters, the central star density and the density at
  the base of the crust. If bare, quark stars possess
  ultra-high electric fields on the order of $10^{18}$ to
  $10^{19}$~V/cm. These features render the properties of quark stars
  more multifaceted than those of neutron stars and may allow one to
  observationally distinguish quark stars from neutron stars.
\end{abstract}


\firstsection 

\section{Introduction}

\begin{figure}[htb]
\begin{center}
 \includegraphics[width=7.0cm,angle=0]{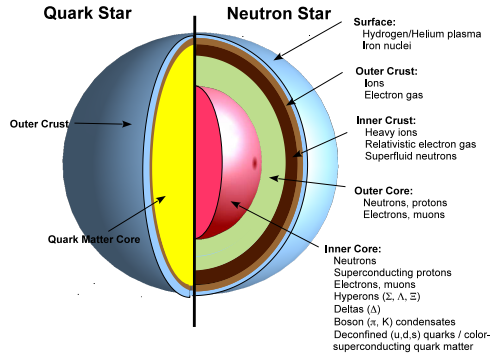} 
 \caption{Schematic structures of quark stars and neutron stars.}
\label{fig:nsss}
\end{center}
\end{figure}

The theoretical possibility that quark matter made of up, down and
strange quarks (so-called strange quark matter (\cite[Farhi \& Jaffe
1984]{farhi84:a})) may be more stable than ordinary nuclear matter has
been pointed out by \cite{bodmer71:a}, \cite{terazawa79:a}, and
\cite{witten84:a}. This so-called strange matter hypothesis
constitutes one of the most startling possibilities regarding the
behavior of superdense matter, which, if true, would have implications
of fundamental importance for cosmology, the early universe, its
evolution to the present day, and astrophysical compact objects such
as neutron stars and white dwarfs (see \cite[Alcock \& Farhi
1986]{alcock86:a}, \cite[Alcock \& Olinto 1988]{alcock88:a},
\cite[Aarhus 1991]{aarhus91:proc}, \cite[Weber 1999]{weber99:book},
\cite[Madsen 1999]{madsen98:b}, \cite[Glendenning 2000]{glen97:book},
\cite[Weber 2005]{weber05:a}, \cite[Page \& Reddy
2006]{page06:review}, 
\cite[Sagert et al. 2006]{sagert06:a},
and references therein). The properties of quark
stars are compared with those of neutron stars in
Table~\ref{tab:comparison} and Fig.\ \ref{fig:nsss}.  Even to the
present day there is no sound scientific basis on which one can either
confirm or reject the hypothesis so that it is a serious possibility
of fundamental significance for various (astro) physical phenomena.

The multifaceted properties of these
\begin{table}
  \begin{center}
    \caption{Theoretical properties of quark stars and neutron stars
      compared.}
  \label{tab:comparison}
  \begin{tabular}{|l|l|}\hline
{\bf Quark Stars}  & {\bf Neutron Stars}  \\ \hline
Made entirely of deconfined up, down, strange      &Nucleons, hyperons, boson condensates, \\ 
quarks, and electrons                              &deconfined quarks, electrons, and muons \\ \hline
Quarks ought to be color superconducting           &Superfluid neutrons \\
                                                   &Superconducting protons \\ \hline
Energy per baryon $\lsim 930$~MeV                  &Energy per baryon $> 930$~MeV \\ \hline
Self-bound ($M \propto R^3$)                       &Bound by gravity \\ \hline
Maximum mass $\sim 2 \, \msun$                     &Same \\ \hline
No minimum mass                                    &$\sim 0.1 \, \msun$ \\ \hline
Radii $R\lsim 10- 12 $~km                          &$R \gsim 10 - 12$~km \\ \hline
Baryon number $B \lsim 10^{57}$                     &$10^{56}  \lsim B \lsim 10^{57}$ \\ \hline
Electric surface fields $\sim 10^{18}$ to $\sim 10^{19}$~V/cm     &Absent   \\ \hline
Can be bare (pure quark stars) or enveloped       &Not possible \\ 
in thin nuclear crusts (mass $10^{-5}\, \msun$)    &Always possess nuclear crusts    \\ \hline
Density of crust is less than neutron drip        &Density of crust above neutron drip   \\ 
i.e., posses only outer crusts                    &i.e., posses inner and outer crusts \\ \hline
Form two-parameter stellar sequences              &Form one-parameter stellar sequences \\ \hline
\end{tabular}
\end{center}
\end{table}
objects are reviewed in this paper. Particular emphasis is is put on
stellar properties such as rapid rotation, ultra-high electric surface
fields, and rotational vortex expulsion, which may allow one to
observationally discriminate between quark stars and neutron stars
and--ultimately--prove or disprove the strange quark matter
hypothesis. Futher information on the existence of qark stars may come
from quark novae, hypothetical types of supernovae which could occur
if neutron stars spontaneously collapse to quark stars (\cite[Ouyed et
al. 2002]{ouyed02:a}).

\goodbreak
\section{Quark-Lepton Composition of Quark Stars}\label{sec:qlc}

Quark star matter is composed of the three lightest quark flavor
states (up, down, and strange quarks). Per hypothesis, the energy per
baryon of such matter is lower than the energy per baryon of the most
stable atomic nucleus, $^{56}{\rm Fe}$.  Since stars in their lowest
energy state are electrically charge neutral to very high precision,
any net positive quark charge must be balanced by electrons.  The
concentration of electrons is largest at low densities due to the
finite strange-quark mass, which leads to a deficit of net negative
quark charge. If quark star matter forms a color superconductor
(\cite[Rajagopal \& Wilczek 2001]{rajagopal01:a}, \cite[Alford
2001]{alford01:a}, \cite[Alford et al. 2008]{alford08:a}, and
references therein) in the Color-Flavor-Locked (CFL) phase
the interiors of quarks stars will be rigorously electrically neutral
with no need for electrons, as shown by \cite{rajagopal01:b}.  For
sufficiently large strange quark masses, however, the low density
regime of quark
star matter is rather expected to form other condensation patterns
(e.g.\ 2SC, CFL-$K^0$, CFL-$K^+$, CFL-$\pi^{0,-}$) in which electrons
will be present (\cite[Rajagopal \& Wilczek 2001]{rajagopal01:a},
\cite[Alford 2001]{alford01:a}, \cite[Alford et
al. 2008]{alford08:a}).  The presence of electrons in quark star
matter is crucial for the possible existence of a nuclear crust on
quark stars.  As shown by \cite{alcock86:a,kettner94:b}, and
\cite{alcock88:a}, the electrons, because they are bound to strange
matter by the Coulomb force rather than the strong force, extend
several hundred fermi beyond the surface of the strange star.
Associated with this electron displacement is a electric dipole layer
which can support, out of contact with the surface of the strange
star, a crust of nuclear material, which it polarizes (\cite[Alcock et
al. 1986]{alcock86:a}, \cite[Alcock \& Olinto 1988]{alcock88:a}). The
maximum possible density at the base of the crust (inner crust
density) is determined by neutron drip, which occurs at about
$4.3\times 10^{11}~\gcmt$.  

\section{Bare versus Dressed Quark Stars and Eddington Limit}

A bare quark star differs qualitatively from a neutron star which has
a density at the surface of about 0.1 to $1~\gcmt$. The thickness of
the quark surface is just $\sim 1~\fm$, the length scale of the strong
interaction. The electrons at the surface of a quark star are held to
quark matter electrostatically, and the thickness of the electron
surface is several hundred fermis.  Since neither component, electrons
and quark matter, is held in place gravitationally, the Eddington
limit to the luminosity that a static surface may emit does not apply,
so that bare quark stars may have photon luminosities much greater than
$10^{38}~\ergs$.  It was shown by \cite{usov98:a} that this value
may be exceeded by many orders of magnitude by the luminosity of $e^+
e^-$ pairs produced by the
Coulomb barrier at the surface of a hot strange star. For a surface
temperature of $\sim 10^{11}$~K, the luminosity in the outflowing pair
plasma was calculated to be as high as $\sim 3 \times 10^{51}~\ergs$.
Such an effect may be a good observational signature of bare strange
stars (\cite[Usov 2001a]{usov01:c},
\cite[Usov 2001b]{usov01:b},
\cite[Usov 1998]{usov98:a}, and
\cite[Cheng \& Harko 2003]{cheng03:a}). If the strange star
is enveloped in a nuclear crust, however, which is gravitationally
bound to the strange star, the surface, made of ordinary atomic
matter, would be subject to the Eddington limit. Hence the photon
emissivity of such a ``dressed'' quark star would be the same as for
an ordinary neutron star.  If quark matter at the stellar surface is
in the CFL phase the process of $e^+ e^-$ pair creation at the stellar
quark matter surface may be turned off.
This may be different for the early stages of a very hot CFL quark
star (\cite[Vogt et al. 2004]{vogt03:a}).

\goodbreak
\section{Mass-Radius Relationship of Quark Stars}

\begin{figure}[tb]
\begin{center}
 \includegraphics[width=8.0cm,angle=0]{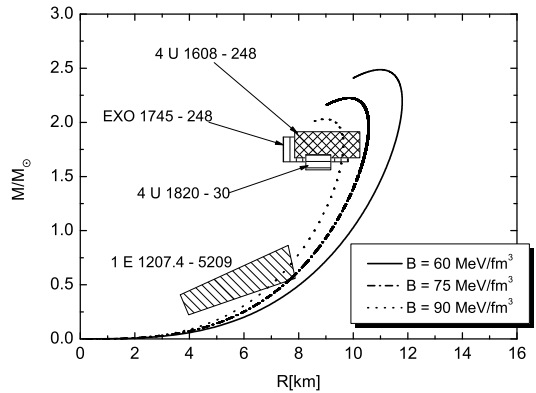} 
 \caption{Mass-radius relationship of bare quark stars (from
   \cite[Orsaria et al. 2011]{orsaria11:a}).}
\label{fig:mr_bare}
\end{center}
\end{figure}

The mass-radius relationship of bare quark stars is shown in Fig.\
\ref{fig:mr_bare}.  In contrast to neutron stars, the radii of
self-bound quark stars decrease the lighter the stars, according to $M
\propto R^3$. The existence of nuclear crusts on quark stars changes
the situation drastically (\cite[Glendenning et al. 1995]{weber95:a},
\cite[Weber 1999]{weber99:book}, and \cite[Weber 2005]{weber05:a}).
Since the crust is bound gravitationally, the mass-radius relationship
of quark stars with crusts is then qualitatively similar to neutron
stars.

In general, quark stars with or without nuclear crusts possess smaller
radii than neutron stars. This feature implies that quark stars posses
smaller mass shedding periods than neutron stars.  Due to the smaller
radii of quarks stars, the complete sequence of such objects--and not
just those close to the mass peak, as it is the case for neutron
stars--can sustain extremely rapid rotation (\cite[Glendenning et
al. 1995]{weber95:a}, \cite[Weber 1999]{weber99:book}, and \cite[Weber
2005]{weber05:a}).  In particular, a strange star with a typical
pulsar mass of around $1.45\,\msun$ can rotate at Kepler (mass
shedding) periods as small as $0.55\lsim{P_{\rm K}}/{\rm msec}\lsim
0.8$
(\cite[Glendenning \& Weber 1992]{glen92:crust}, and \cite[Glendenning
et al. 1995]{weber93:b}). This range is to be compared with ${P_{\rm
    K}} \sim 1~{\rm msec}$ obtained for neutron stars of the same mass
(\cite[Weber 1999]{weber99:book}).

Another novelty of the strange quark matter hypothesis concerns the
existence of a new class of white-dwarf-like objects, referred to as
strange (quark matter) dwarfs (\cite[Glendenning et
al. 1995]{weber93:b}). The mass-radius relationship of the latter may
differs somewhat from the mass-radius relationship of ordinary
white-dwarf, which may be testable in the future. Until
recently, only rather vague tests of the theoretical mass-radius
relation of white dwarfs were possible. This has changed dramatically
because of the availability of new data emerging from the Hipparcos
project (\cite[Provencal 1998]{provencal98:a}). These data allow the
first accurate measurements of white dwarf distances and, as a result,
establishing the mass-radius relation of such objects empirically.

\goodbreak
\section{Pulsar Glitches}
\label{ssec:crust}

Of considerable relevance for the viability of the strange matter
hypothesis is the question of whether strange stars can exhibit
glitches in rotation frequency. From the study performed by
\cite[Glendenning \& Weber (1992)]{glen92:crust} and \cite[Zdunik et
al. (2001)]{zdunik01:a} it is known that the ratio of the crustal
moment of inertia to the total moment of inertia varies between
$10^{-3}$ and $\sim 10^{-5}$.  If the angular momentum of the pulsar
is conserved in a stellar quake one obtains for the change of the
star's frequency ${{\Delta \Omega} / {\Omega}} \sim (10^{-5} -
10^{-3}) f$, where $0 < f < 1$ (\cite[Glendenning \& Weber
1992]{glen92:crust}).  The factor $f$ represents the fraction of the
crustal moment of inertia that is altered in the quake.  Since the
observed glitches have relative frequency changes $\Delta
\Omega/\Omega = (10^{-9} - 10^{-6})$, a change in the crustal moment
of inertia of $f\lsim 0.1$ would cause a giant glitch
(\cite[Glendenning \& Weber 1992]{glen92:crust}). Moreover it turns
out that the observed range of the fractional change in the spin-down
rate, $\dot \Omega$, is consistent with the crust having a small
moment of inertia and the quake involving only a small fraction $f$ of
that.  For this purpose we write $ { {\Delta \dot\Omega} / {\dot\Omega
  } } > (10^{-1}\; {\rm to} \; 10) f$ (\cite[Glendenning \& Weber
1992]{glen92:crust}). This relation yields a small $f$ value, i.e., $f
< (10^{-4} \; {\rm to} \; 10^{-1})$, in agreement with $f\lsim 0.1$
established just above. For these estimates, the measured values of
$(\Delta \Omega / \Omega)/(\Delta\dot\Omega/\dot\Omega) \sim 10^{-6}$
to $10^{-4}$ for Crab and Vela, respectively, have been used.

\goodbreak
\section{Possible Connection to CCOs}

One of the most amazing features of quark stars concerns the possible
existence of ultra-high electric fields on their surfaces, which, for
ordinary quark matter, is around $10^{18}$~V/cm.  If strange matter
forms a color superconductor, as expected for such matter, the
strength of the electric field may increase to values that exceed
$10^{19}$~V/cm. The energy density associated with such huge electric
fields is on the same order of magnitude as the energy density of
strange matter itself, which alters the masses and radii of strange
quark stars at the 15\% and 5\% level, respectively (\cite[Negreiros
et al. 2009]{negreiros09:a}). Such mass increases facilitate the
interpretation of massive compact stars, with masses of around $2 \,
M_\odot$, as strange quark stars (see also \cite[Rodrigues et
  al. 2011]{rodrigues11:a}).

The electrons at the surface of a quark star are not necessarily in a
fixed position but may rotate with respect to the quark matter star
(\cite[Negreiros et al. 2010]{negreiros10:a}). In this event magnetic
fields can be generated which, for moderate effective rotational
frequencies between the electron layer and the stellar body, agree
with the magnetic fields inferred for several Compact Central Objects
(CCOs). These objects could thus be interpreted as quark stars whose
electron atmospheres rotate at frequencies that are moderately
different ($\sim 10$~Hz) from the rotational frequency of the quark
star itself.

Last but not least, we mention that the electron surface layer may be
strongly affected by the magnetic field of a quark star in such a way
that the electron layer performs vortex hydrodynamical oscillations
(\cite[Xu et al. 2012]{xu12:a}). The frequency spectrum of these
oscillations has been derived in analytic form by \cite{xu12:a}. If the
thermal X-ray spectra of quark stars are modulated by vortex
hydrodynamical oscillations, the thermal spectra of compact stars,
foremost central compact objects (CCOs) and X-ray dim isolated neutron
stars (XDINSs), could be used to verify the existence of these
vibrational modes observationally. The central compact object 1E
1207.4-5209 appears particularly interesting in this context, since
its absorption features at 0.7 keV and 1.4 keV can be comfortably
explained in the framework of the hydro-cyclotron oscillation model
(\cite[Xu et al. 2012]{xu12:a}).

A study which looks at the thermal evolution of CCOs is presently
being carried out by \cite{shuhua12:a}. Preliminary results indicate
that the observed temperatures of CCOs can be well reproduced if one
assumes that these objects are small quark matter objects with radii
less than around 3~km.

\goodbreak
\section{Possible Connection to SGRs, AXPs,  and XDINs}

If quarks stars are made of color superconducting quark matter rather
than normal non-superconducting quark matter.  If rotating,
superconducting quark stars ought to be threaded with rotational
vortex lines, within which the star's interior magnetic field is at
least partially confined.  The vortices (and thus magnetic flux) would
be expelled from the star during stellar spin-down, leading to
magnetic reconnection at the surface of the star and the prolific
production of thermal energy. \cite{niebergal10:a} have shown
that this energy release can re-heat quark stars to exceptionally high
temperatures, such as observed for Soft Gamma Repeaters (SGRs),
Anomalous X-Ray pulsars (AXPs), and X-ray dim isolated neutron stars
(XDINs). Moreover, numerical investigations of the temperature
evolution, spin-down rate, and magnetic field behavior of such
superconducting quark stars suggest that SGRs, AXPs, and XDINs may be
linked ancestrally (\cite[Niebergal et al 2010]{niebergal10:a}).

\acknowledgments M. Orsaria thanks CONICET for financial
support. H. Rodrigues thanks CAPES for financial support under
contract number BEX 6379/10-9. F. Weber is supported by the National
Science Foundation (USA) under Grant PHY-0854699. S.-H. Yang is
supported by the China Scholarship Council (CSC) and by NFSC under Grant
No.\ 11147170.

\end{document}